\documentclass[sigconf]{acmart}

\usepackage{array}
\usepackage{tabularx}
\usepackage{subcaption}
\usepackage{graphicx}
\usepackage[most]{tcolorbox}
\usepackage[utf8]{inputenc}
\usepackage{textgreek}

\tcbset{
  rqbox/.style={
    colback=gray!10!white,   
    colframe=pink!70,       
    fonttitle=\bfseries,
    coltitle=black,
    boxrule=0.8pt,
    arc=3mm,
    left=4mm,
    right=4mm,
    top=2mm,
    bottom=2mm,
    before skip=10pt,
    after skip=10pt
  }
}

\copyrightyear{2026}
\acmYear{2026}
\setcopyright{cc}
\setcctype{by}
\acmConference[MSR '26]{23rd International Conference on Mining Software Repositories}{April 13--14, 2026}{Rio de Janeiro, Brazil}
\acmBooktitle{23rd International Conference on Mining Software Repositories (MSR '26), April 13--14, 2026, Rio de Janeiro, Brazil}
\acmPrice{}
\acmDOI{10.1145/3793302.3793376}
\acmISBN{979-8-4007-2474-9/2026/04}

\begin{document}

\title{Analyzing GitHub Issues and Pull Requests in nf-core Pipelines: Insights into nf-core Pipeline Repositories}

\author{Khairul Alam}
\email{kha060@usask.ca}
\authornotemark[1]
\affiliation{%
  \institution{University of Saskatchewan}
  \city{Saskatoon}
  \state{Saskatchewan}
  \country{Canada}
}

\author{Banani Roy}
\email{banani.roy@usask.ca}
\affiliation{%
  \institution{University of Saskatchewan}
  \city{Saskatoon}
  \state{Saskatchewan}
  \country{Canada}
}

\renewcommand{\shortauthors}{Trovato et al.}

\begin{abstract}
Scientific Workflow Systems (SWSs) such as Nextflow have become essential software frameworks for conducting reproducible, scalable, and portable computational analyses in data-intensive fields like genomics, transcriptomics, and proteomics. Building on Nextflow, the nf-core community curates standardized, peer-reviewed pipelines that follow strict testing, documentation, and governance guidelines. Despite its widespread adoption, little is known about the challenges users face in developing and maintaining these pipelines. This paper presents an empirical study of 25,173 issues and pull requests from these pipelines to uncover recurring challenges, management practices, and perceived difficulties. Using BERTopic modeling, we identify 13 key challenges, including pipeline development and integration, bug fixing, integrating genomic data, managing CI configurations, and handling version updates. We then examine issue-resolution dynamics, showing that 89.38\% of issues and pull requests are eventually closed, with half resolved within 3 days. Statistical analysis reveals that the presence of labels (large effect, $\mathit{d} = 0.94$) and code snippets (medium effect, $\mathit{d} = 0.50$) significantly improves the likelihood of resolution. Further analysis reveals that tool development and repository maintenance poses the most significant challenges, followed by testing pipelines and CI configurations, and debugging containerized pipelines. Overall, this study provides actionable insights into the collaborative development and maintenance of nf-core pipelines, highlighting opportunities to enhance their usability, sustainability, and reproducibility.
\end{abstract}



\maketitle

\section{Introduction}
The rapid growth of high-throughput sequencing and other omics (e.g., genomics, transcriptomics, and metagenomics) technologies has led to an explosion of large, heterogeneous, and complex bioinformatics datasets \cite{yamada2021interpretation}. Analyzing these data requires chaining together multiple computational tools (modules) with diverse dependencies, formats, and parameter configurations. To address these challenges, scientific workflow systems (SWSs) such as Nextflow \cite{di2017nextflow}, Snakemake \cite{koster2012snakemake}, and Galaxy \cite{galaxy2022galaxy} have emerged as indispensable software frameworks, enabling researchers to design, execute, and share multi-step computational analyses with improved reproducibility, scalability, and portability \cite{deelman2009workflows, ewels2020nf}.

One of the most widely adopted SWSs is Nextflow \cite{di2017nextflow}, which provides a domain-specific language for defining pipelines and integrates seamlessly with package managers (e.g., Conda) and containerization technologies (e.g., Docker, Singularity). Nextflow also supports deployment on various infrastructures, including local workstations, high-performance computing (HPC) clusters, and cloud platforms. This flexibility allows developers to prototype pipelines locally and subsequently scale them to large computational resources with minimal configuration changes.

Building upon Nextflow, the nf-core community \cite{nfcore_website} has developed a large ecosystem of standardized, peer-reviewed, and well-documented pipelines \cite{ewels2020nf}. These pipelines\footnote{https://nf-co.re/pipelines/} support diverse domains ranging from transcriptomics and genomics to proteomics and imaging. Hosted on GitHub, these pipelines promote community-driven development, with each release versioned and assigned a DOI via Zenodo for reproducibility. Through best-practice guidelines, comprehensive documentation, continuous integration, and containerization, nf-core has become a leading platform for reproducible bioinformatics workflows \cite{ewels2020nf}.
While similar initiatives, such as Snakemake Workflows \cite{snakemake_workflows_github}, Galaxy ToolShed \cite{blankenberg2014dissemination}, ENCODE \cite{sloan2016encode}, and WorkflowHub \cite{da2020workflowhub}, also promote workflow sharing and reuse, nf-core stands out for its unified governance and automation-driven approach to quality assurance. Understanding what users discuss within nf-core pipeline repositories is therefore vital: it can expose recurring technical pain points (e.g., dependency management, environment setup, and infrastructure compatibility), highlight common feature requests and enhancement trends, and illuminate collaboration dynamics among domain scientists, bioinformaticians, and developers that collectively shape workflow sustainability and evolution.

From a development standpoint, nf-core offers a standardized pipeline template that enforces coding style and guidelines, reducing entry barriers and promoting code consistency. Automated synchronization propagates template updates, while governance, code reviews, and integration with Bioconda/conda-forge support collaborative and sustainable development. Despite these strengths, maintaining nf-core pipelines remains challenging. Contributors must manage dependencies across diverse environments, resolve synchronization conflicts, update documentation, and ensure rigorous testing. These tasks are coordinated primarily through GitHub issues and pull requests, which facilitate bug reports, feature requests, and discussions. While prior studies have examined collaboration and maintenance through GitHub artifacts \cite{alam2025drives, yang2023users, li2021understanding, alam2025empirical}, systematic analyses of the types of questions and issues raised within nf-core's bioinformatics workflows are still lacking.

In this paper, we address this gap through an empirical study of GitHub issues and pull requests from nf-core pipeline repositories. Using BERTopic modeling, we identify 13 key challenge areas, including pipeline development and integration, bug fixing, genome data integration, CI management, and version updates, and examine how developers manage and resolve them. Our analysis shows that 89.38\% of issues and pull requests are resolved, with 50.74\% closed within three days. Across 125 active repositories (12 are archived out of 137 total as of August 22, 2025), three have no open issues or pull requests, and two have none closed. About 4.64\% of issues and pull requests remain entirely ignored. We also observe that 92.74\% (115) of repositories utilize labeling, while 90.32\% (112) assign issues and pull requests to specific contributors. Beyond the default labels, we find three additional frequently used labels among the top ten, namely \emph{WIP, DSL2}, and \emph{docs}. Among closed cases, 60.88\% are self-closed, whereas only 17.13\% involve assigned contributors. Furthermore, our analysis indicates that issues and pull requests with labels and code blocks are more likely to be closed, suggesting that providing appropriate labels and essential code snippets can enhance the clarity and resolution of reported issues and pull requests. We make the following contributions in this paper:

\textbf{1. }We conduct an empirical study to systematically investigate issues and pull requests in open-source analysis pipelines within the nf-core ecosystem.  By analyzing 25,173 issues and pull requests from 125 nf-core pipeline repositories, our study provides the workflow community with a comprehensive dataset that facilitates a deeper understanding of collaboration, maintenance, and development practices.

\textbf{2. }We categorize user-reported challenges into a taxonomy of 13 topics, offering insights into recurring problems and development practices. This categorization aims to support developers in better understanding user challenges, thereby guiding improvements in pipeline usability, functionality, and overall maintenance.

\textbf{3. }Our findings highlight the value of good practices in managing issues and pull requests. We recommend actively using GitHub features (e.g., labeling, assigning) for organization and accountability, and encourage contributors to provide clear, well-formatted descriptions with code snippets to improve clarity and resolution efficiency.



\section{Background and Related Work}
\label{nf-core-background}
\subsection{Scientific Workflow Systems and Pipelines}
An SWS is a specialized software platform that enables researchers to design, execute, and manage complex computational experiments often referred to as scientific workflows \cite{liu2015survey}. These systems automate repetitive tasks, integrate diverse tools, and ensure reproducibility by tracking data provenance. Examples include Galaxy \cite{galaxy2022galaxy}, Nextflow \cite{di2017nextflow}, Snakemake \cite{koster2012snakemake}, Pegasus \cite{deelman2015pegasus}, and Taverna \cite{oinn2004taverna}.

A pipeline is a concrete instance of a scientific workflow \cite{barker2007scientific}, representing a sequence of data processing steps tailored to a specific research task, such as RNA-seq analysis in bioinformatics. Pipelines can be graph-based or script-based, and developers can use domain-specific language (DSL) \cite{fowler2010domain}, common workflow language (CWL) \cite{amstutz2016common}, workflow description language (WDL), or yet another workflow language (YAWL) \cite{van2005yawl} to develop pipelines. Community repositories such as nf-core curate standardized pipelines. Examples include rnaseq \cite{nfcore_rnaseq_3_21_0}, sarek \cite{nfcore_sarek_3_5_1} and methylseq \cite{nfcore_methylseq_4_1_0}. In short, the SWS is the platform, while the pipeline is the recipe. For example, Nextflow provides an engine for running an RNA-seq pipeline, ensuring scalability, reproducibility, and efficient execution.

\subsection{nf-core Pipeline Repositories}
The nf-core project develops reproducible pipelines using Nextflow \cite{ewels2020nf}. Supported by many leading organizations such as \emph{Seqera, SciLifeLab, the Center for Genomic Regulation, AWS, and Microsoft Azure}, it ensures high-quality, portable analyses across diverse environments. Through shared maintenance, peer review, and centralized management, it fosters collaboration and enhances reuse. As of August 22, 2025, it hosts 125 active pipelines and over 1,600 modules \cite{nfcore-modules}, exemplifying a sustainable, community-led model for bioinformatics research \cite{nfcore-pipelines}.
\begin{figure*}[htbp]
    \centering
    \includegraphics[width=0.8\textwidth, height=0.4\textheight]{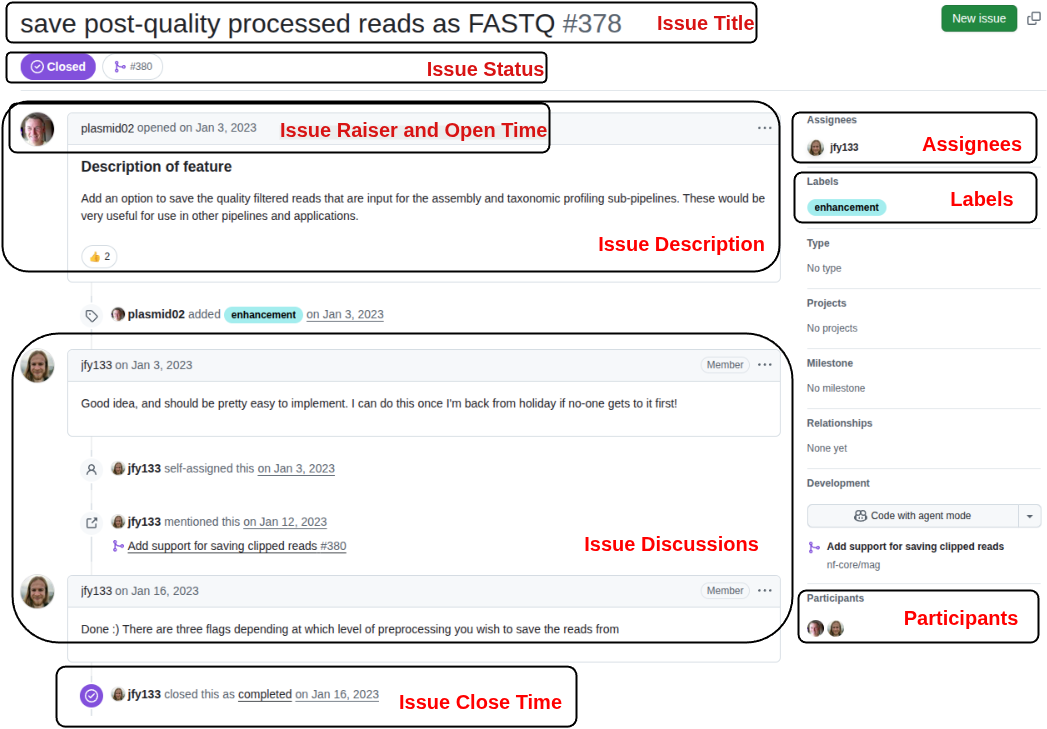}
    \vspace{-.5em}
    \caption{An illustrative example of an issue raised in the nf-core/mag \cite{krakau2022nf} pipeline on GitHub. }
    \label{fig:nf-core-workflow-sample}
\end{figure*}
\subsection{Tracking Issues and Pull Requests in GitHub}
Software systems are not always reliable \cite{gundersen2018reproducible}, and maintaining open-source repositories presents ongoing challenges. To facilitate collaboration, GitHub introduced \emph{Pull Requests} in 2008 \cite{pr-tracker-action} to propose, review, and merge code changes and \emph{Issues} in 2009 \cite{github-issue-tracker} for reporting bugs, discussing features, and asking questions. Figure~\ref{fig:nf-core-workflow-sample} illustrates an issue from the \emph{nf-core/mag}\footnote{\url{https://github.com/nf-core/mag/issues/378}} pipeline designed for the assembly, binning, and annotation of metagenomes. In this example, the individual who created the issue, referred to as the issue raiser, provides a title and a description that specify the reported concern. Any GitHub user with access to the repository may contribute to the discussion through comments, which often provide clarifications, additional details, or potential solutions. For management purposes, issues are supported by two fundamental mechanisms: \textit{labels} and \textit{assignees}. Labels can be created by contributors with write access to the repository and applied or removed by those with triage permissions, thereby enabling classification and prioritization. Assignees can be designated by users with write access, ensuring that responsibility for resolution is explicitly assigned. All actors involved in the lifecycle of an issue, including the raiser, commenters, and assignees, are collectively regarded as participants. Both issues and pull requests share features such as titles, descriptions, labels, and threaded discussions. Pull requests, however, add code-specific functionality, commit and diff views, inline reviews, automated CI/CD checks, and merge strategies.

\subsection{Mining GitHub Issues and Pull Requests}
GitHub repositories provide a rich data source for understanding developer challenges, collaboration, and software evolution. Prior work has mined issues and pull requests to study collaboration patterns \cite{tsay2014influence}, pull-based development models, and software evolution dynamics \cite{gousios2014exploratory}. Domain-specific studies have examined repositories for identifying developers' pain points for deep learning frameworks \cite{han2020programmers}, Quantum Software Engineering \cite{li2021understanding}, AI systems \cite{yang2023users}, and scientific workflow systems \cite{alam2025empirical}. Our study contributes to this literature by systematically analyzing over 25K issues and pull requests from nf-core, providing the first large-scale taxonomy of user concerns in pipeline repositories.

\subsection{Topic Modeling}
Topic modeling is a statistical technique for finding latent themes in large text collections, facilitating organization and understanding of unstructured content \cite{jelodar2019latent}. It has been widely applied in document clustering, information retrieval, recommendation systems, and software engineering research, particularly in mining software repositories and source code analysis \cite{asuncion2010software, bagheri2014adm, gethers2010using}. Traditional models such as LDA and its variants (e.g., sLDA, CTM, RTM, DTM) \cite{blei2003latent} have been widely used but suffer from fixed topic numbers, lack of contextual understanding, and limited scalability. They are prone to generating incoherent topics, particularly in the presence of noisy or extensive datasets, and their performance is susceptible to hyperparameter settings, requiring extensive experimentation \cite{alam2025empirical}. In contrast, BERTopic \cite{grootendorst2022bertopic} leverages transformer-based embeddings, and produces more coherent, context-aware, and interpretable topics. Its ability to handle technical, short, and noisy texts such as GitHub data makes it particularly suitable for this study.

\subsection{Topic Analysis of GitHub Data}
Topic modeling with GitHub data involves analyzing issues, pull requests, and comments to identify key topics. This technique can uncover trends, common issues, or popular features within software development communities. Researchers apply algorithms like LDA, BERTopic, and others to categorize the vast amount of unstructured text data on GitHub to understand challenges in various domains, such as QSE \cite{li2021understanding}, cryptography APIs \cite{nadi2016jumping}, open-source AI repositories \cite{yang2023users}, deep learning frameworks \cite{han2020programmers}, desktop web apps \cite{scoccia2021challenges}, scientific workflow systems development \cite{alam2025empirical} and others \cite{dhasade2020towards, jokhio2021mining, wang2019does}. In addition, the mining of GitHub repositories has become a cornerstone in empirical research for uncovering collaboration patterns \cite{gousios2014exploratory}, studying software evolution \cite{kalliamvakou2014promises}, and characterizing issues and pull requests management practices \cite{tsay2014influence}.

\section{Study Design}
\label{nf-core-methodology}
\subsection{Data Collection}
As of August 22, 2025, the nf-core framework comprises 137 pipelines, of which 12 have been archived. Since archived pipelines are no longer actively maintained, they are excluded from our analysis, resulting in a final dataset of 125 active pipelines. Using the GitHub REST API, we collect 25,586 issues and pull requests across these pipelines. However, as many pull requests are directly linked to existing issues, treating them as independent records could introduce redundancy. To address this, we employ the GitHub Search API to identify and exclude 410 pull requests. Thus, in total, we have 25,176 issues and pull requests. Among all pipelines, nf-core/sarek \cite{nfcore_sarek_3_5_1} has the highest number of issues and pull requests (1,947), followed by nf-core/rnaseq \cite{nfcore_rnaseq_3_21_0} with 1,583. To enrich our dataset with repository-level context, we retrieve metadata for each repository under study. The collected metadata includes key attributes such as the number of contributors, stars, forks, and watchers of each repository. It is important to note that nf-core pipelines are continuously evolving. For this study, we collect issues and pull requests spanning from March 2018, coinciding with the establishment of nf-core, through August 22, 2025. We share the data collection scripts in the replication package \cite{random_2025_17307554} for transparency and reproducibility.

\subsection{Pre-processing Issues and Pull Requests}
To enable reliable topic modeling of GitHub data, we perform a systematic preprocessing to clean and normalize the textual data. Specifically, we concatenate the title and body of each issue or pull request into a single document to preserve sufficient contextual information for topic extraction. We then remove code snippets using regular expressions, as such blocks often contain syntax elements irrelevant to semantic modeling. Similarly, we eliminate URLs, which typically link to external resources, and clean out residual HTML tags. We further normalize the corpus by removing special characters and punctuation. To address high-frequency but semantically uninformative words, we apply stopword removal using the NLTK stopword list \cite{hardeniya2016natural}, and supplement it by filtering out project-specific terms such as \emph{nf-core} and \emph {nextflow} to prevent these frequently occurring system names from dominating the topic space. Finally, we apply lemmatization \cite{balakrishnan2014stemming} using the \texttt{spaCy en\_core\_web\_sm} model \cite{vasiliev2020natural} to convert words into their base forms, thus improving topic coherence and reducing lexical variation. This preprocessing ensures that the data retains its semantic richness while reducing noise, making it suitable for subsequent topic modeling and analysis. After preprocessing, we identify three samples with empty titles and bodies (these originally contained only URLs in the title and empty body). As they provide no meaningful information, we exclude them from the analysis. Consequently, the final dataset consists of \textbf{25,173} issues and pull requests.

\subsection{Identifying Topics in nf-core Pipelines}
BERTopic begins by transforming input documents into numerical representations that capture their semantic content. Among several approaches, \emph{Sentence Transformers} \cite{sentence_transformers_huggingface} has emerged as a state-of-the-art technique for producing sentence and document embeddings, widely recognized for its ability to encode semantic similarity across short and noisy texts. It provides a diverse set of pre-trained models on the \emph{Hugging Face Model Hub} \cite{huggingface_models}, including the family of \emph{all-*} models that were trained on over one billion training pairs. For our study, we adopt \emph{all-mpnet-base-v2}, a fine-tuned version of Microsoft’s MPNet that has consistently achieved high performance on semantic textual similarity benchmarks such as  Semantic Textual Similarity  and the Massive Text Embedding Benchmark. This model generates 768-dimensional embeddings optimized for sentence-level semantic similarity and has been shown to perform well in clustering tasks. Its ability to capture nuanced linguistic features makes it especially well-suited for analyzing short, technical, and domain-specific texts, such as GitHub data.

\textbf{Determining topic modeling parameters: }In BERTopic, the \emph{nr\_topics} parameter allows users to control the number of topics by merging similar ones after their initial creation. However, prior work \cite{alam2025empirical, li2021understanding} has emphasized the value of relying on clustering algorithms to produce a more natural topic structure. To enhance clustering quality, we first reduce the dimensionality of the document embeddings, as high-dimensional data can hinder performance. For this step, we employ \emph{UMAP} \citep{mcinnes2018umap}, which effectively preserves local semantic structure while projecting embeddings into a lower-dimensional space. We then apply \emph{HDBSCAN} \citep{mcinnes2017hdbscan}, a density-based algorithm particularly well suited to \emph{UMAP} projections. \emph{HDBSCAN} is capable of identifying noise points and does not force every document into a cluster, making it especially effective for GitHub data.

For BERTopic modeling, we systematically tune \emph{UMAP} and \emph{HDBSCAN} parameters to balance topic granularity with semantic coherence. To evaluate the resulting topics, we employ the topic coherence score \cite{roder2015exploring}, a widely used metric that measures the semantic similarity among the top words in a topic. This measure helps distinguish between topics that are semantically interpretable and those that are merely artifacts of statistical inference. For UMAP, we vary \emph{n\_neighbors} (15–50) to balance local semantic similarity with a meaningful global structure, and adjust \emph{n\_components} (5–30) to retain sufficient semantic information while enabling efficient clustering. We use \emph{cosine distance}, which aligns well with sentence-transformer embeddings. For HDBSCAN, we experiment with \emph{min\_cluster\_size} values between 30 and 200, adopting \emph{Euclidean distance} as it is well suited to UMAP-reduced spaces. To improve topic representation, we apply \emph{CountVectorizer} with \emph{unigram} and \emph{bigram} models, consistent with prior studies \cite{alam2025empirical, li2021understanding, abdellatif2020challenges}.

We achieve the highest topic coherence score of 0.904 for our dataset using \emph{n\_neighbors} = 30, \emph{n\_components} = 15, and \emph{min\_cluster\_size} = 150. With this configuration, BERTopic produces 13 distinct topics. However, BERTopic introduces a special \emph{-1} topic to aggregate documents that do not align well with any of the discovered topics. Such cases typically occur when documents contain excessive noise (e.g., stopwords or irrelevant content) or lack strong connections to the main topics the model extracts. Consequently, the \emph{-1} topic acts as a \emph{catch-all} cluster for outlier documents. Since our dataset underwent extensive preprocessing to remove noisy and irrelevant elements prior to topic modeling, our main topic began with the label \emph{-1}. The preprocessing and topic generation scripts are available in our replication package \citep{random_2025_17307554}. A detailed analysis of the resulting topics is presented in the Section~\ref{nf-core-results}.

\smallskip
\noindent The \textbf{replication package} is available in our online appendix \cite{random_2025_17307554}.

\section{Results} 
\label{nf-core-results}
In this section, we analyze the issues and pull requests from the nf-core pipelines. Our experiments are designed to address the following three research questions:
\begin{itemize}
    \item \textbf{RQ1: }What topics do users raise and discuss in the issues and pull requests of nf-core pipelines?
    \item \textbf{RQ2: }How are issues and pull requests managed and addressed in nf-core pipelines, and how do different features influence their resolution?
    \item \textbf{RQ3: }To what extent do the practitioners find the development activities relevant to the topics they discuss in the issues and pull requests challenging?
\end{itemize}
\subsection{RQ1: What topics do users raise and discuss?}
\textbf{Motivation: }Understanding the topics raised in issues and pull requests is essential to identifying the challenges and needs in nf-core pipeline development and maintenance. Since these pipelines are community-driven and widely used in bioinformatics, the issues and pull requests often reflect recurring technical difficulties, usability concerns, feature requests, and collaborative practices. Importantly, the development of nf-core pipelines differs substantially from that of traditional software systems: pipelines must integrate heterogeneous bioinformatics tools, manage complex data dependencies, and ensure reproducibility across diverse computational environments (e.g., containers, HPC clusters, cloud platforms). These requirements introduce domain-specific challenges, such as compatibility with reference datasets, versioned tools, and workflow execution engines, which are less prevalent in general-purpose software engineering \cite{di2017nextflow, arvanitou2021software, gruening2019recommendations}. Therefore, the challenges developers and users face in scientific workflow repositories such as nf-core may differ from those encountered in traditional software projects. The motivation behind this \textbf{RQ1} is to reveal the \emph{challenges, needs, and recurring concerns} of the nf-core pipeline repositories.

\textbf{Approach: }To identify the key topics, we apply BERTopic modeling to our datasets, as outlined in Section~\ref{nf-core-methodology}. The modeling process yields 13 distinct topics along with the distribution of co-occurring words for each. We then manually assign meaningful, contextually accurate labels to the identified topics. 
Following established practices in prior work~\cite{alam2025empirical, li2021understanding, openja2020analysis, yang2016security, bagherzadeh2019going, scoccia2021challenges}, the first author, having over five years of experience in nf-core pipelines and more than a decade of professional software development experience, proposes initial candidate labels. These are based on topic keywords and a manual review of at least 30 randomly selected representative issues and pull requests per topic. The proposed labels are subsequently refined through collaborative discussions with the second author, who has over nine years of experience in bioinformatics-related pipeline research, and an additional expert with more than two decades of expertise in empirical software engineering research. All disagreements were resolved through structured discussions. Through this iterative process, we finalize a set of labels that accurately capture the underlying semantics and contextual nuances of the identified topics.

\textbf{Results: }Table~\ref{tab:nf-core-topics} summarizes the identified topics, their associated keywords, and a brief description of each topic. Below, we provide a more detailed discussion of each topic.
\begin{table*}[htbp]
\footnotesize
    \centering
    \caption{Topics derived from issues and pull requests of nf-core pipelines on GitHub}
    \label{tab:nf-core-topics}
      \begin{tabularx}{\linewidth}{r|>{\raggedright\arraybackslash}p{3.6cm}|>{\raggedright\arraybackslash}p{5.8cm}|>{\raggedright\arraybackslash}p{6.9cm}}
        \toprule
        SL. & Topic & Keywords & Description \\
        \midrule
        1 & Pipeline Development and Integration & pipeline, add test, documentation update, branch, description, new tool, file, include, change reason & Covers initial pipeline development tasks such as creating tools, adding tests, preparing documentation, and managing branches. \\
        \hline
        2 & Template Synchronization and Automated Update Management & nfcoretool template, conflict branch, make resolve, relevant update, automate attempt, instruction information, update template, make new, update pipeline & Captures automated nf-core template updates, the resolution of branch conflicts, and contributor efforts to apply template changes consistently across pipelines. \\
             \hline
        3 &  Debugging Execution Failures & debug, version, singularity, description bug, command use, executor, fail, output, information, pipeline & Addresses debugging and execution failures in containerized environments, command usage, and executor behavior. \\
             \hline
        4 & Maintaining Tools, Tests, and Documentation & tool, documentation update, add test, fix, repository make, nfcoretestdataset repository, follow pipeline, update output & Highlights the challenges in keeping tools, tests, and documentation consistent and up to date across nf-core pipelines.\\
             \hline
        5 & Bug Fixing & fix bug, add test, reason, make documentation, update learn, nfcoretestdataset, necessary, description change &  Captures the recurring problems of fixing bugs, adding tests in nf-core pipelines. \\
             \hline
        6 &  Coordinating Contributions and Maintaining Pipeline Quality & contribute, documentation update, tool, add test, new, pipeline learn, make branch, fix, appropriate delete & Reflects users' difficulties in managing collaborative contributions, updating documentation, and ensuring high-quality tools and tests. \\
             \hline
        7 & Genome Data Integration and Testing & genome, add, sequence, description feature, alignment, run, tool, test, include, variant & Covers difficulties in integrating genome data, running alignment tools, and testing workflows for sequence and variant analyses. \\
             \hline
        8 & Modules Maintenance & module, documentation, reason, make, pipeline, description change, parameter, conda update, feature add & Covers challenges in updating modules, managing parameters, and keeping documentation aligned with feature changes. \\
             \hline
       9 & Subworkflow Development and Integration & subworkflow, new tool, add test, documentation update, branch, nfcoreraredisease, include, description feature, fix & Discuss users' difficulties in developing subworkflows, integrating them, and ensuring consistent documentation, tests, and fixes. \\
            \hline
        10 & Tool Development and Repository Maintenance &  make, new tool, contribute, documentation update,  add test, branch pipeline, relevant common, appropriate delete. & Highlights challenges in developing new tools, managing contributions, updating documentation, and maintaining repositories. \\
             \hline
        11 & Testing Pipelines and Managing CI Configurations & ci, aws, need, pipeline, config, datum, description feature, test dataset, run & Addresses challenges in running pipelines in CI environments, managing configurations, and maintaining test datasets. \\
             \hline
        12 & Reporting and Quality Control Visualization & multiqc report, description feature, table, plot, stat, add, qc, module, read, general & Reflects challenges to generate, interpret, and extend reporting features such as MultiQC modules, plots, and quality control tables. \\
             \hline
        13 & Version Updates and Repository Maintenance & changelog, commit, master, branch, template sync, merging, update version, chores, final, automate & Highlights challenges in version updates, commits, and repository maintenance, including changelogs, branching, and automation.\\
        \bottomrule
    \end{tabularx}
    \vspace{-1em}
\end{table*}

\textbf{1. Pipeline Development and Integration: }This is the most dominant topic users face when working with nf-core pipelines, focusing on the foundational tasks of initiating and integrating new workflows in the nf-core ecosystem. For example, in the \emph{hlatyping} pipeline, an issue, \emph{nf-core pipelines should provide structured output information}, proposed extending nf-core metadata to not only describe inputs but also provide structured output information, helping humans and automated systems anticipate outputs, and route results. Developers frequently scaffold pipelines using standardized templates, add files and test stubs, and ensure compliance with nf-core community guidelines. Keywords such as \emph{new tool, add test, pipeline}, and \emph{documentation update} highlight a strong emphasis on code quality and adherence to standards, with many issues and pull requests focusing on early development tasks.
\newline
\textbf{2. Template Sync. and Automated Update Management: }This is the second most frequent challenge, highlighting the need to keep pipelines synchronized with the latest nf-core template. Issues and pull requests in this topic involve propagating new conventions and updates, resolving merge conflicts, and ensuring community standards. For instance, the \emph{dualrnaseq} pipeline issue titled as \emph{Pipeline needs updating to the latest template version} required fixing the \emph{TEMPLATE} branch and creating a sync pull request to resolve conflicts. Keywords such as \emph{conflict branch}, \emph{make resolve}, and \emph{automate attempt} illustrate the hybrid nature of this process, where automation drives standardization but human intervention remains essential.
\newline
\textbf{3. Debugging Execution Failures: }This topic captures the practical challenges users encounter when debugging nf-core pipelines during execution. Failures often result from incorrect command usage, incompatible software versions, or backend issues. For instance, the issue \emph{singularity fails to pull the hic-1.3.0 container} reports an aborted run due to image retrieval errors, pointing to problems with the container registry or singularity conversion. Keywords such as \emph{singularity}, \emph{executor}, and \emph{debug} highlight recurring difficulties in containerized environments, where runtime errors and system-specific setups complicate troubleshooting.
\newline
\textbf{4. Maintaining Tools, Tests, and Documentation: }Users encounter challenges in sustaining workflow quality and consistency as tools, datasets, and documentation evolve. Typical issues include incomplete documentation, failing tests, and inconsistent input and outputs, often reflected in keywords like \emph{add test, fix}, and \emph{update output}. For example, the pull request \emph{Fix input schema to allow direct input again} addresses input schema validation by enabling direct input in multiple formats (e.g., CSV, FASTQ files).
\newline
 \textbf{5. Bug Fixing: }Users often encounter challenges to ensure pipeline reliability by resolving bugs, adding tests, and maintaining representative datasets (\emph{e.g., fix bug, add test, nfcoretestdataset}). For example, the pull request \emph{Bugfix MALT error retries} addresses a memory-related issue by enabling the pipeline to retry when Java exits with code 1 due to heap space exhaustion, rather than aborting immediately.
\newline
\textbf{6. Coordinating Contributions and Maintaining Pipeline Quality: }Users often struggle to balance multiple tasks when contributing, such as updating documentation, fixing issues, and adding tests. Keywords like \emph{contribute}, \emph{make branch}, and \emph{appropriate delete} reflect recurring coordination challenges across branches and the cleanup of outdated content. For example, the pull request \emph{Revert multiqc workaround} restores original functionality by removing a temporary fix after a patch release resolved the underlying issue. 
\newline
\textbf{7. Genome Data Integration and Testing: }Users often encounter challenges in incorporating genome data and related analyses into pipelines. Keywords such as \emph{genome, sequence, alignment}, and \emph{variant} point to difficulties with data integration and tools execution. For example, the issue \emph{Handle several metagenome samples not individually but synergistically}, aims to enable the pipeline to pool multiple metagenomic samples together (rather than processing each in isolation) to improve assembly and binning results. This topic underscores the difficulty of maintaining a reliable pipeline for genomic analyses.
\newline
\textbf{8. Module Maintenance:  }Users face challenges in updating modules, adjusting parameters, and keeping their documentation consistent. Keywords such as \emph{module, parameter}, and \emph{feature add} indicate technical work tied to implementing new functionality or refining existing modules. An example of this topic is \emph{Add metagenomics module}, which aims to integrate metagenomic classification of unmapped reads into the \emph{eager} pipeline.
\newline
\textbf{9. Subworkflow Development and Integration: } Users often struggle to incorporate subworkflows into existing nf-core pipelines. Keywords such as \emph{subworkflow}, \emph{new tool}, and \emph{fix} highlight the technical effort needed to align contributions while maintaining compatibility. For example, the \emph{deepmodeloptim} pipeline issue \emph{Adapt subworkflow to the nf-core structure} demonstrates the need to conform to nf-core’s modular design and standards.
\newline
\textbf{10. Tool Development and Repository Maintenance: }Users frequently encounter challenges when creating new tools and maintaining the repositories. Keywords such as \emph{make} and \emph{new tool} reflect to the complexity of initiating tool development, while \emph{add test}, \emph{branch pipeline}, and \emph{appropriate delete} indicate difficulties in ensuring correctness, managing branches, and removing outdated files. For example, the pull request \emph{Add Kraken2 processes to build database and filter read} illustrates these efforts.
\newline
\textbf{11. Testing Pipelines and Managing CI Configurations: }Users often face challenges testing pipelines in continuous integration (CI) environments, particularly when using cloud resources like AWS. Keywords such as \emph{ci}, \emph{aws}, and \emph{config} highlight difficulties in configuring environments to ensure reliable test execution across systems. Issues such as \emph{CI: add gpu tag test job} illustrate efforts to balance fast, automated CI testing with the complexity of scientific pipelines that depend on large datasets and diverse configurations.
\newline
\textbf{12. Reporting and Quality Control Visualization: }This topic focuses on reporting and visualization challenges of nf-core pipelines. Keywords such as \emph{multiqc report}, \emph{table}, \emph{plot}, and \emph{stat} highlight difficulties in presenting quality control data clearly and consistently. Common issues involve adding new QC modules, extending report features, or fixing display inconsistencies. For example, \emph{Add functionality to turn read QC on/off} illustrates such efforts.
\newline
\textbf{13. Version Updates and Repository Maintenance: }Users often struggle to keep nf-core pipelines aligned with evolving versions and repository structures. Issues in this category involve tasks such as updating changelogs, managing commits, and coordinating version releases. Keywords like \emph{commit}, \emph{update version}, and \emph{changelog} reflect the effort required for documentation and version management, while \emph{branch}, \emph{merging}, and \emph{final} highlight the complexities of synchronizing updates across branches.

\begin{tcolorbox}[rqbox,title={RQ1 Summary}]
In \textbf{RQ1}, we identify 13 distinct topics through BERTopic modeling and manual validation. These topics encompass areas such as pipeline development and integration, template synchronization, debugging execution failures, tool and documentation maintenance, contribution  coordination, genome data integration, module and parameter updates, subworkflow development, CI configuration, reporting/visualization, and version management. The taxonomy demonstrates that user discussions in nf-core repositories span the entire workflow lifecycle.
\end{tcolorbox}

\subsection{RQ2: Management and resolution of issues and pull requests in nf-core pipelines.}
\textbf{Motivation: }
In nf-core pipelines, GitHub issues and pull requests serve as the primary channels for reporting bugs, requesting features, and contributing improvements. They reflect both the technical challenges users encounter and the collaborative processes used to resolve them. Understanding how these artifacts are managed is essential, as their resolution directly affects pipeline reliability, usability, and sustainability. As highlighted in \textbf{RQ1}, users raise issues and pull requests for diverse reasons, and ignoring these challenges can prevent users from successfully using these pipelines. While some cases, such as installing required packages, can be easily addressed, others, like tool development, CI management, and debugging, require coordinated effort and management.

While prior studies have explored issues and pull requests management in general open-source projects, pipeline repositories like nf-core pose unique, domain-specific challenges that may shape how issues are prioritized, resolved, or neglected. Investigating how these processes are managed and how features such as labeling, assignment, or code snippets influence resolution can provide important insights into the socio-technical dynamics of collaborative pipeline maintenance. Thus, \textbf{RQ2} seeks to uncover the practices and strategies that sustain effective issue/pull request management in nf-core, offering lessons for both the nf-core community and the broader ecosystem of reproducible pipeline development.

\begin{figure}[htbp]
    \centering
    \begin{subfigure}[b]{0.32\columnwidth}
        \centering
        \includegraphics[width=\textwidth, height=0.20\textheight]{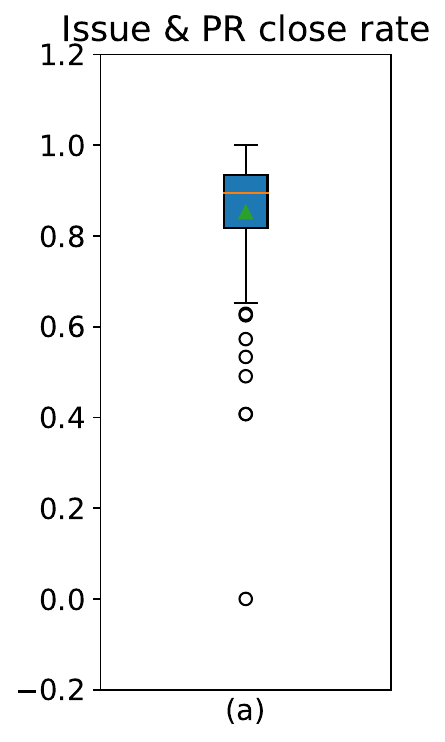} 
        \label{fig:1a}
    \end{subfigure}
    \hfill
    \begin{subfigure}[b]{0.32\columnwidth}
        \centering
        \includegraphics[width=\textwidth, height=0.20\textheight]{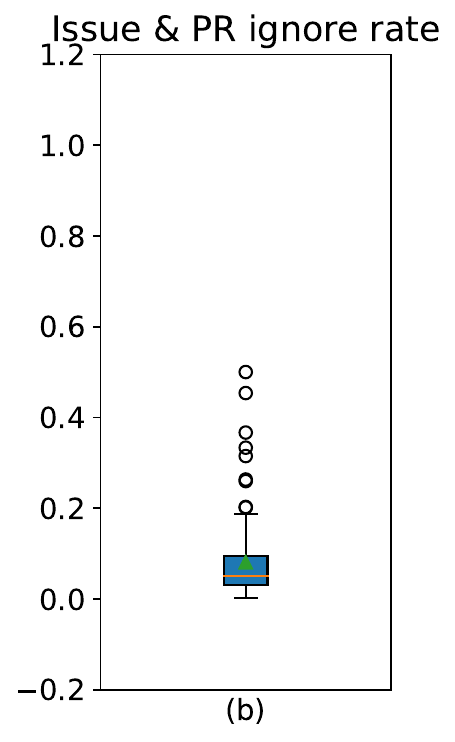} 
        \label{fig:1b}
    \end{subfigure}
    \hfill
    \begin{subfigure}[b]{0.32\columnwidth}
        \centering
        \includegraphics[width=\textwidth, height=0.20\textheight]{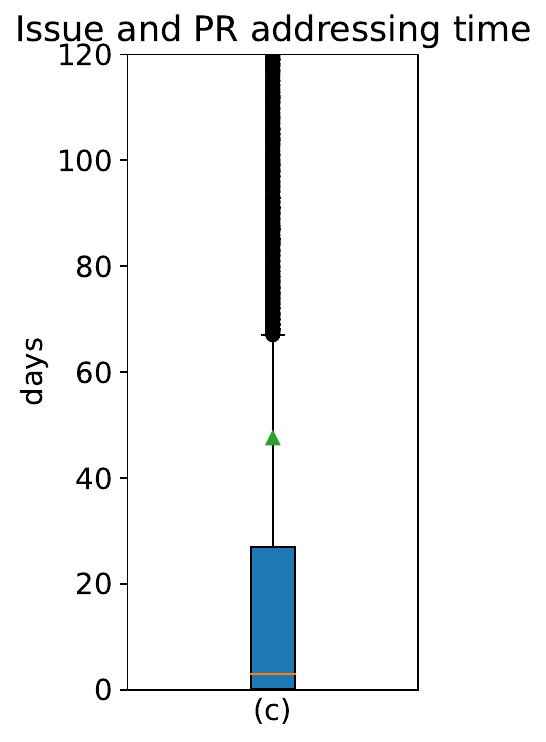}
        \label{fig:1c}
    \end{subfigure}
    \vspace{-1.8em}
  \caption{The distribution of (a) closed issues and pull requests rate, (b) ignored issues and pull requests rate, and (c) addressing time.}
  \label{fig:issue-comparison}
\end{figure}

\textbf{How are the issues and pull requests addressed?}
We examine how issues and pull requests are managed in our dataset. Among 25,173 issues and pull requests, 22,500 (89.38\%) are closed, and 2,673 (10.62\%) remained open, with an overall average of 201.38 entries per repository. Figure~\ref{fig:issue-comparison}(a) illustrates the distribution of the closed issues and pull requests rate (CIR) across repositories. Notably, 1.61\% of repositories have achieved a CIR of 1.0, indicating that all reported issues and pull requests have been resolved. Moreover, repositories with a CIR exceeding 0.9 exhibit an average of 259.13 issues and pull requests, suggesting that repositories with higher closure rates also tend to manage a larger volume of entries. Overall, 48.39\% of repositories demonstrate a CIR greater than 0.9, highlighting that nearly half of the repositories maintain a consistently high resolution rate. 

We find that 60.88\% of closed issues and pull requests are \emph{self-closed}, meaning that the individuals who reported them were able to resolve their concerns, often through subsequent discussion. In addition, we identify another type of case referred to
as \emph{ignored issues and pull requests}. These are cases that only the individuals who raised them are interested in, and receive no response from other users. These ignored cases account for 4.64 \% of all issues and pull requests. We also deﬁne the \emph{ignored issues and pull requests rate} as the ratio of ignored cases for each repository. This metric provides insight into the extent of active engagement by repository maintainers in addressing cases. The distribution of this rate across repositories is presented in Figure~\ref{fig:issue-comparison}(b).

We also analyze issues and pull requests addressing time, defined as the duration between creation and closure. Among 22,500 closed cases, the mean addressing time is 47.51 days with a standard deviation of 141.67 days, while the median is 2.89 days, indicating most are resolved quickly. However, a subset remains unresolved for extended periods. The maximum addressing time observed exceeds four years, highlighting significant variability in resolution dynamics. Figure~\ref{fig:issue-comparison}(c) shows the distribution of the addressing time.

To examine factors influencing issue and pull requests resolution time, we analyze two repository-level features following prior work \cite{yang2023users}: (1) the number of contributors and (2) the total number of issues and pull requests. We hypothesize that a larger contributor base may reduce addressing time. A negative Spearman correlation of –0.198 (p = 0.0283) between contributors and median closing time indicates that repositories with more contributors resolve issues slightly faster. Although this effect is statistically significant, the practical impact remains small. We further hypothesize that repositories with more issues and pull requests would exhibit longer addressing times, as a larger workload might burden maintainers. Contrary to this expectation, the results reveal a Spearman correlation of –0.218 (p = 0.0157) between the number of issues and pull requests and the median closing time. This negative correlation suggests that repositories handling a greater volume of issues/pull requests actually resolve them somewhat faster, though again, the effect size is modest.

\begin{figure}[htbp]
  \centering
    \includegraphics[width=0.7\linewidth, height=0.2\textwidth]{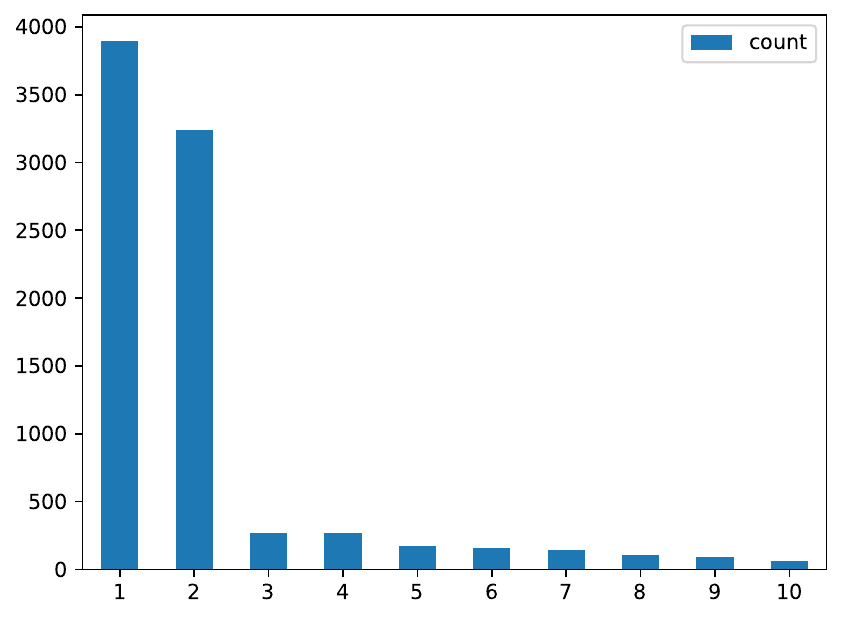}
  \caption{The distribution of the top 10 most frequently used labels. The numbers represent the labels, where \textbf{1:}  \emph{enhancement}, \textbf{2:} \emph{bug}, \textbf{3:} \emph{question}, \textbf{4:} \emph{documentation}, \textbf{5:} \emph{WIP}, \textbf{6:} \emph{good
first issue}, \textbf{7:} \emph{feature-request}, \textbf{8:} \emph{help wanted},
\textbf{9:} \emph{DSL2}, \textbf{10:} \emph{docs}.}
  \label{fig:labels-rank}
\end{figure}
\vspace{-1.0em}
\textbf{How are the issues and pull requests managed? }GitHub offers several mechanisms to support maintainers in managing issues and pull requests. Following prior work \cite{yang2023users}, we focus on two commonly used strategies: (1) \textit{labeling}, where issues and pull requests are tagged with labels, and (2) \textit{assigning}, where issues and pull requests are allocated to specific individuals for resolution. In the remainder of this section, we use the terms \emph{labeling} and \emph{assigning} to refer to these two practices.

We observe that 33.23\% (8,366) of issues and  pull requests are tagged with labels, while 17.13\% are assigned to specific contributors. This indicates that a large portion of issues and  pull requests do not make use of GitHub’s built-in management features. To further examine the use of \emph{labels}, we note that GitHub provides nine default options: \emph{bug, documentation, duplicate, enhancement, help wanted, invalid, question, wontfix, and good first issue}. Actively managed projects often go beyond these defaults by defining additional labels (e.g., DSL2). Across all pipeline repositories, we identify 154 distinct labels (including the defaults). We then calculate the frequency of each label, with the ranked distribution presented in Figure~\ref{fig:labels-rank}.
\newline
\textbf{Relationships between different features and the closure of issues and pull requests: }We further examine how different characteristics of issues and  pull requests relate to their likelihood of closure. Understanding these relationships offers practical insights for repository maintainers and contributors on improving the management and resolution of issues and  pull requests. Specifically, we focus on the following features:
\newline
\textbf{1. has-label:} Indicates whether an issue/pull request has a label. A labeled entry usually means that it has been noticed and read by the maintainers, thereby increasing the likelihood of being addressed.
\newline
\textbf{2. has-assignee:} An assigned issue/pull request typically indicates clear responsibility, which may increase the likelihood of timely and successful resolution.
\newline
\textbf{3. title-length:}  Refers to the length of an issue or pull request title. Informative, longer titles provide more context, helping maintainers understand and resolve items more quickly.
\newline
\textbf{4. body-length: }Detailed descriptions give maintainers essential context, aiding triage and resolution, and are often encouraged by project guidelines.
\newline
\textbf{5. has-code: }Indicates if the issue or pull request includes code blocks, which help maintainers reproduce and resolve issues more effectively.
\newline
\textbf{6. has-URL: }Indicates if the issue or pull request contains URLs, which provide helpful context through links to related issues, documentation, or external resources.
\newline
 \textbf{7. num-contributors: }Intuitively, repositories with a larger contributor base have greater collective capacity, which may increase the chances of timely issue/pull request resolution.

\textbf{Feature Measurement and Statistical Test: }We compute the length of the issue/pull request title and body. Using regular expressions, we then identify whether an issue/pull request description contains code blocks or URLs. The number of contributors is obtained by querying the GitHub API to fetch metadata about the repository. Finally, we apply the Wilcoxon rank-sum test \cite{wilcoxon1945individual} to assess the statistical significance of the differences in feature distributions between open and closed issues and pull requests.
\vspace{-1.em}
\begin{table}[htbp]
\footnotesize
\centering
\caption{The significance of the differences between the feature values of open and closed issues and pull requests (Wilcoxon rank-sum test and Cohen's effect sizes $\mathit{d}$).}
\label{tab:feature_significance}
\begin{tabular}{lcc}
\hline
\textbf{Feature} & \textbf{P-value} & \textbf{Cohen's $\mathit{d}$} \\
\hline
has-label        & $<$0.01 & 0.94 (Large) \\
has-assignee     & $<$0.01 & 0.09 (Negligible) \\
title-length     & $<$0.01 & 0.28 (Small) \\
body-length      & $<$0.01 & 0.001 (Negligible) \\
has-code         & $<$0.01 & 0.50 (Medium) \\
has-URLs         & $<$0.01 & 0.37 (Small) \\
num-contributors & $<$0.01 & 0.14 (Negligible) \\
\hline
\end{tabular}
\end{table}
\vspace{-1em}

We also estimate Cohen’s effect size ($\mathit{d}$) to quantify the magnitude of differences between open and closed issues and pull requests. Following established guidelines, values of $\mathit{d} < 0.2$ are regarded as negligible, $0.2 \leq |\mathit{d}| < 0.5$ as small, $0.5 \leq |\mathit{d}| < 0.8$ as medium, and $|\mathit{d}| \geq 0.8$ as large. Table~\ref{tab:feature_significance} reports the results of the Wilcoxon rank-sum test and corresponding Cohen’s $\mathit{d}$ effect sizes when comparing open and closed issues and pull requests across different features. All features exhibit statistically significant differences ($p < 0.01$). However, the effect sizes indicate that the practical significance varies notably across features.

\begin{itemize}
    \item \textbf{has-label }shows a large effect size ($\mathit{d} = 0.94$), suggesting that closed issues/pull requests are far more likely to be labeled compared to open issues and pull requests.
    \item \textbf{has-code }presents a medium effect size ($\mathit{d} = 0.50$), implying that the presence of code snippets differs meaningfully between open and closed issues and pull requests.
    \item \textbf{title-length} ($\mathit{d} = 0.28$) and \textbf{has-URL }($\mathit{d} = 0.37$) exhibit small effect sizes, showing modest differences.
    \item \textbf{has-assignee }($\mathit{d} = 0.09$), \textbf{body-length }($\mathit{d} = 0.001$), and \textbf{num-contributors }($\mathit{d} = 0.14$) have negligible effect sizes, indicating that although statistically significant, the differences in these features are not practically meaningful.
\end{itemize}
Overall, the results highlight that \emph{labeling} and the inclusion of \emph{code snippets} are the most distinguishing features between open and closed issues and pull requests, while differences in \emph{assignee presence}, \emph{body length}, and \emph{contributor count} are less impactful.
\begin{tcolorbox}[rqbox,title={RQ2 Summary}]
In \textbf{RQ2}, we find that 89.38\% are closed, with nearly half resolved within three days. Feature analysis shows that repositories making active use of GitHub's management mechanisms, labeling (used in 33.23\% of cases) and assignment (17.13\%), achieve higher closure rates. Statistical tests reveal that labeling (large effect,$\mathit{d} = 0.94$) and including code snippets (medium effect, $\mathit{d} = 0.5$) are the strongest predictors of successful resolution, while title length and URLs show smaller effects.
\end{tcolorbox}

\subsection{RQ3: Perceived relevance of development activities to challenging topics.}
\textbf{Motivation: }Building on the topics, our next step is to examine the difficulty of resolving challenges within each category. Not all issues carry the same weight: routine tasks like version updates or reporting and quality control visualization are relatively straightforward, whereas debugging containerized workflows, integrating genomic data, or maintaining pipeline quality often require deeper expertise, extensive coordination, and more time. Assessing perceived difficulty is important for several reasons. First, it highlights which categories impose the greatest cognitive and technical burden, guiding improvements in tooling, documentation, and contributor support. Second, it sheds light on the socio-technical complexity of maintaining the pipelines, where technical tasks must be balanced with collaborative processes. Moreover, it helps the community prioritize resources and streamline practices around the most demanding areas, thereby enhancing both the sustainability and usability of nf-core pipelines.
\newline
\textbf{Approach: }We evaluate the difficulty of each topic using two complementary metrics following established practices in prior studies \citep{alam2025empirical, li2021understanding, scoccia2021challenges}. The details of these metrics are outlined below:
\newline
\textbf{1. The percentage of unresolved issues and pull requests of a topic (\% w/o solutions)}  reflects how many are unresolved within that particular topic area. For each topic identified in \textbf{RQ1}, we calculate the proportion of issues and pull requests that have not been closed. Closing typically indicates that a developer has addressed the problem or implemented the requested features and thus considers it resolved. Consequently, topics with a higher proportion of open issues and pull requests can be considered more difficult.
\newline    
\textbf{2. The median time for issues and pull requests to be resolved (Median time to resolve (hr)).} We calculate the median resolution time for issues and pull requests based on the interval between their creation and closing times. This metric applies only to closed issues and pull requests, as open ones lack a closing timestamp. A longer resolution time indicates greater difficulty in addressing the issue or implementing the requested changes.

\begin{table}[htbp]
    \centering
    \footnotesize
    \caption{Unresolved \% and Median Time for Different Topics}
    \label{tab:unresolved_issues_median_time_nf-core}
    \vspace{-.5em}
    \begin{tabularx}{\linewidth}{X|>{\raggedleft\arraybackslash}p{1.2cm}|>{\raggedleft\arraybackslash}p{1.0cm}}
        \toprule
        \textbf{Topic Name} & \textbf{Unresolved (\%)} & \textbf{Median Time (h)} \\
        \midrule
        Pipeline Development and Integration & 10.68 & 69.78 \\
        Template Sync. and Automated Update Management & 10.63 & 65.89 \\
         Debugging Execution Failures & 10.72 & 70.24 \\
        Maintaining Tools, Tests, and Documentation & 9.18 & 62.59 \\
         Bug Fixing & 5.53 & 69.46 \\
       Coordinating Cont. \& Maintaining Pipeline Quality & 9.02 & 71.49 \\
       Genome Data Integration and Testing  & 11.04 & 72.57 \\
        Modules Maintenance & 7.96 & 74.11 \\
        Subworkflow Development and Integration & 11.72 & 66.17 \\
        Tool Development and Repository Maintenance & 20.28 & 36.83 \\
        Testing Pipelines and Managing CI Configurations & 12.17 & 36.44 \\
        Reporting and Quality Control Visualization & 9.44 & 29.37 \\
        Version Updates and Repository Maintenance &12.43 & 55.88 \\
        \bottomrule
    \end{tabularx}
\end{table}

\textbf{Results: }
Table \ref{tab:unresolved_issues_median_time_nf-core} presents the proportion of unresolved issues and pull requests and the median time required to close them across the topics identified in \textbf{RQ1}. The results reveal substantial variation in both persistence and resolution speed, highlighting differing levels of complexity across topic categories.

The highest unresolved rate is observed in \emph{Tool Development and Repository Maintenance} (20.28\%), indicating that tasks related to introducing new tools, refactoring repositories, and maintaining cross-pipeline dependencies remain particularly challenging. Despite a relatively short median resolution time (36.83 h), the high proportion of unresolved cases suggests that while simpler maintenance tasks are closed quickly, complex tool-related efforts often stall, likely due to dependency conflicts, required domain expertise, or cross-pipeline dependencies. Similarly, \emph{Testing Pipelines and Managing CI Configurations} (12.17\%, 36.44 h) shows a notable difficulty level. Issues and pull requests in this category involve cloud-based execution, large datasets, or hardware constraints, making debugging and reproducibility time-consuming. \emph{Genome Data Integration and Testing} (11.04\%, 72.57 h) and \emph{Debugging Execution Failures} (10.72\%, 70.24 h) also exhibit long resolution times, reflecting the computational and data-specific challenges of running containerized workflows and managing diverse sequencing datasets.

By contrast, topics like \emph{Reporting and Quality Control Visualization} (9.44\%, 29.37 h) and \emph{Version Updates and Repo. Maintenance} (12.43\%, 55.88 h) are resolved more efficiently, benefiting from automated scripts and standardized documentation. Likewise, \emph{Maintaining Tools, Tests, and Documentation} (9.18\%, 62.59 h) shows moderate difficulty, suggesting that structured templates and community guidelines help streamline updates.

Overall, issues and pull requests tied to integration, debugging, and infrastructure management remain the most persistent and time-intensive, whereas documentation, visualization, and versioning tasks are comparatively well-managed. These findings underscore the need for stronger automation, clearer contributor guidelines, and enhanced CI diagnostics to mitigate the high effort associated with complex technical tasks in nf-core pipeline maintenance.

\begin{tcolorbox}[rqbox,title={RQ3 Summary}]
Analysis of \textbf{RQ3} shows that the most complex areas are \textit{Tool Development and Repository Maintenance} (20.28\%, 36.83 h), \textit{Testing Pipelines and Managing CI Configurations} (12.17\%, 36.44 h), and \textit{Genome Data Integration and Testing} (11.04\%, 72.57 h), reflecting the high technical and coordination effort required for integration, debugging, and infrastructure management. Similarly, \textit{Debugging Execution Failures} (10.72\%, 70.24 h) and \textit{Subworkflow Development and Integration} (11.72\%, 66.17 h) exhibit prolonged resolution times, underscoring their complexity.
\end{tcolorbox}

\section{Discussion}
\label{nf-core-discussion}

\begin{figure}[htbp]
\vspace{-1.em}
  \centering
    \includegraphics[width=0.8\linewidth, height=0.2\textwidth]{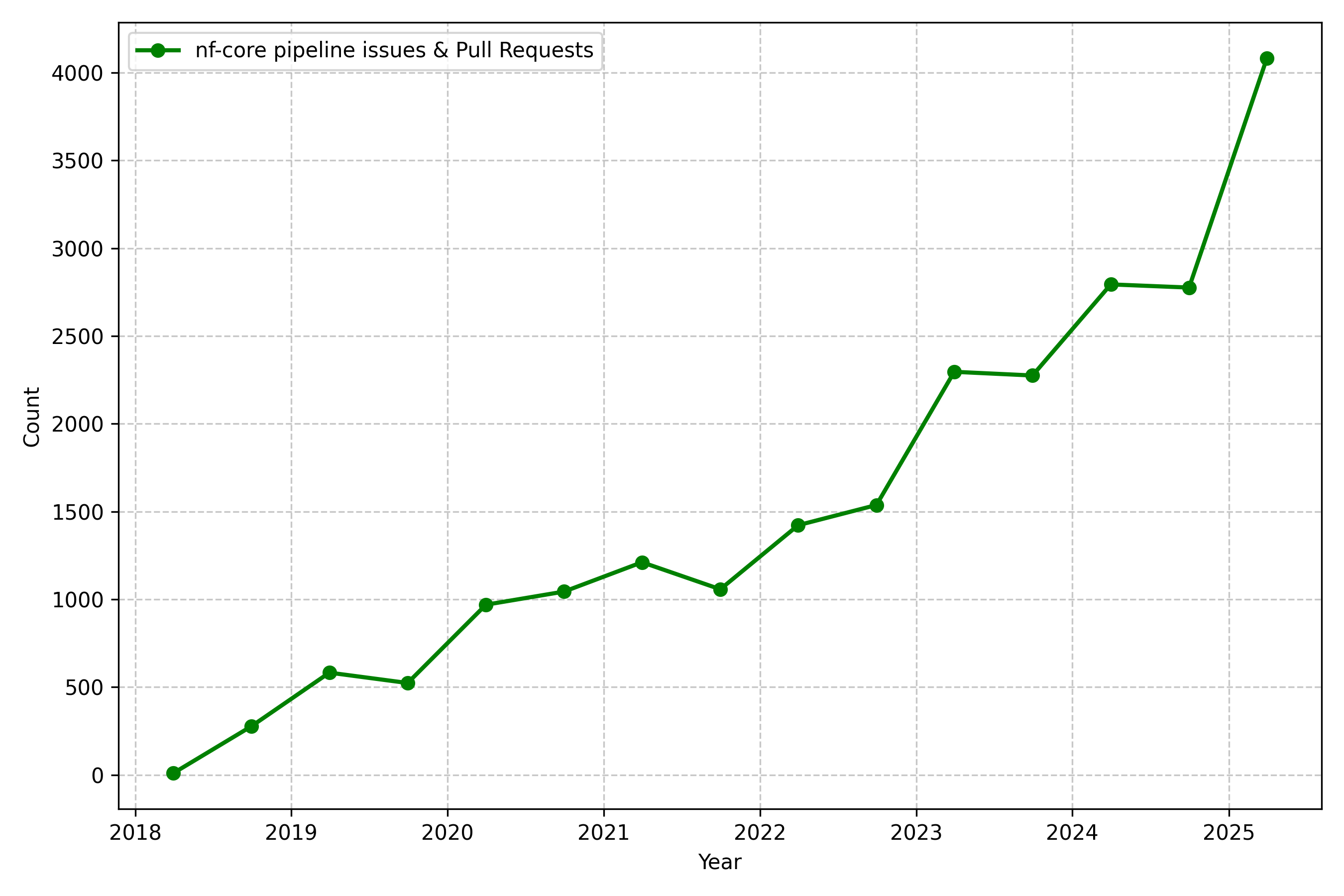}
  \caption{Yearly distribution of issues and pull requests}
  \label{fig:issues_prs_distribution}
  \vspace{-1.5em}
\end{figure}

In this study, we analyze 25,173 issues and pull requests from 125 nf-core pipelines, averaging 201.38 entries per repository. Among the pipelines, \emph{sarek} (1,947), \emph{rnaseq} (1,583), and \emph{eager} (1,057) contain the largest number of issues and pull requests. On average, repositories received 55.45 stars, 36.86 forks, and 151.55 watchers, with \textit{rnaseq} emerging as the most popular (1,093 stars, 793 forks) and \textit{mitodetect} attracting the most watchers (230). As shown in Figure~\ref{fig:issues_prs_distribution}, issues and pull requests activity have steadily increased, particularly after 2022. 
This growth underscores nf-core's evolution into a collaborative ecosystem, where active maintenance, and contribution practices demonstrate the platform's central role in promoting standardized, reusable, and community-driven bioinformatics pipelines.

Our findings reveal key insights into the nf-core pipeline development. First, the topic taxonomy \textbf{(RQ1)} shows that discussions span the entire pipeline lifecycle, highlighting nf-core's dual role as both a software engineering and bioinformatics community. Unlike typical open-source projects, nf-core discussions emphasize reproducibility, container compatibility, and reference data management. Second, the analysis of issue and pull request management \textbf{(RQ2)} shows high closure rates and quick resolutions, but also reliance on informal or implicit coordination, suggesting the need for scalable governance as the ecosystem grows. Third, difficulty assessments \textbf{(RQ3)} identify tool development, CI configuration, and integration-heavy tasks as the most demanding, requiring stronger automation and contributor guidance. Collectively, these findings portray nf-core pipelines as a mature and sustainable collaborative ecosystem. Its evolution mirrors broader trends in open-source scientific infrastructure, where automation and community governance reduce maintenance burdens. Future work could extend this study by analyzing contributor trajectories, comparing nf-core pipelines to other OSS ecosystems, and exploring AI-based triage or summarization tools to further ease community workload.

\section{Implications}
\label{nf-core-implication}
\textbf{For practitioners and maintainers: }The study underscores the value of structured management practices. The strong association between labeling and resolution success suggests that systematic triaging can greatly enhance efficiency. Repositories should adopt clear labeling schemes, automated triage, and mechanisms for contributor assignment. Given that tool development and CI configuration are especially challenging, maintainers should also offer better documentation, onboarding resources, and reusable templates to ease contributor participation.
\newline
\textbf{For contributors: }Detailed and well-structured reports are key to effective collaboration. Issues and pull requests with longer descriptions and embedded code snippets show higher closure rates, indicating that contributors can enhance responsiveness by providing sufficient context and referencing related documentation.
\newline
\textbf{For the broader community: }The nf-core ecosystem shows how a domain-specific community can achieve high responsiveness and sustainability by balancing automation with human expertise. However, reliance on informal coordination may present challenges as the community scales.  Strengthening governance through clearer contribution guidelines, structured review processes, and community training will be essential for sustaining long-term growth.
\newline
\textbf{For researchers: }Our taxonomy of topics provides a foundation for comparative studies across domains (e.g., bioinformatics, AI, data science), while the identified difficulties highlight opportunities to develop automated tools for CI management, dependency resolution, and reproducibility assurance.

\section{Threats to Validity}
\label{nf-core-threats-to-validity}
\textbf{Internal validity: }Our analysis relies on BERTopic to cluster GitHub issues and pull requests. Different modeling choices, such as alternative clustering methods, preprocessing steps, or embedding models, could yield different topics. To ensure quality, we combine systematic preprocessing (e.g., lemmatization, stopword removal, improved vectorization) with manual validation. Three experienced researchers collaboratively review and label the generated topics, resolving disagreements through consensus. While this mitigates subjectivity, topic labeling involves some interpretive bias.
\newline
\textbf{Construct Validity:} Our study relies on data from GitHub, the most widely used platform for managing issues and pull requests in nf-core pipeline development. Although GitHub offers rich and structured insights, it may not fully capture all practitioner challenges, particularly those discussed in external sources. We assess topic difficulty using two indicators: unresolved issues and pull requests, and median resolution time. While these measures are informative, they may not fully capture all dimensions of difficulty, potentially threatening construct validity. Moreover, several confounding factors, including issue type, pipeline complexity and maturity, community engagement, and developer experience, may influence resolution outcomes. Additional indicators, such as the number of comments, labels, review comments, and review cycles, could provide further nuance. Nevertheless, the selected metrics are consistent with those employed in prior related studies \cite{alam2025empirical, li2021understanding, abdellatif2020challenges}.
\newline
\textbf{External Validity: }We focus on nf-core, a well-established ecosystem of Nextflow-based pipelines. While this makes our findings representative of large, domain-specific workflow repositories, they may not generalize to smaller or less structured open-source communities. Similarly, results drawn from bioinformatics workflows may not fully extend to other scientific domains such as cheminformatics, climate science, or machine learning.
\newline
\textbf{Conclusion Validity: }Statistical associations should be interpreted cautiously. Although we use non-parametric tests and effect size measures to ensure robust comparisons, the presence of confounding variables (e.g., repository size, contributor experience) may still influence results. However, rigorous testing and transparent reporting support the reliability and validity of our conclusions.
\vspace{-1.2em}
\section{Conclusion}
\label{conclusion}
This paper provides the first empirical analysis of 25,173 GitHub issues and pull requests across 125 active nf-core pipeline repositories to understand the challenges users face during the development and maintenance of these pipelines. Through BERTopic modeling, we identify 13 topics that span the entire pipeline lifecycle, from pipeline scaffolding and module updates to debugging, continuous integration, and quality control reporting. Our results show that 89.38\% of issues and pull requests are resolved, with half closed within three days, demonstrating an active and responsive community. Statistical analysis further reveals that labeling and inclusion of code snippets strongly influence resolution success, underscoring the importance of structured communication and transparent triage. We also find that tool development, CI configuration, and genome data integration are the most demanding areas, requiring deeper expertise and coordination. In contrast, documentation, and version management tasks are relatively more straightforward. 

In the future, we intend to conduct surveys and interviews to uncover the reasons behind the lack of widespread adoption of issues and pull requests management features in nf-core pipeline repositories and to explore ways to increase their usage. We also plan to investigate the topics of discussion surrounding the development of pipelines using different SWSs, such as Galaxy and Snakemake.

\section*{Acknowledgement}
This research is supported in part by the Natural Sciences and Engineering Research Council of Canada (NSERC), and by the industry-stream NSERC CREATE in Software Analytics Research (SOAR)
\bibliographystyle{ACM-Reference-Format}
\bibliography{sample-base}

\end{document}